\documentclass[preprint,showpacs,preprintnumbers,amsmath,amssymb]{revtex4}

\usepackage{graphicx}
\usepackage{dcolumn}
\usepackage{bm}

\let\vp\varphi
\let\p\partial

\let\non\nonumber

\allowdisplaybreaks

\begin{document}


\title{Separable unsteady nonparallel flow stability problems}

\author{Georgy I. Burde}
 \email{georg@bgu.ac.il}
\affiliation{ Jacob Blaustein Institute for Desert Research,
Ben-Gurion University, Sede-Boker Campus, 84990, Israel}

\author{Alexander Zhalij}
 \email{zhaliy@imath.kiev.ua}
\affiliation{ Institute of Mathematics of the Academy of Sciences
of Ukraine, Tereshchenkivska Street 3, 01601 Kyiv-4, Ukraine}

\begin{abstract}

The so-called 'direct' approach to separation of variables in
linear PDEs is applied to the hydrodynamic stability problem.
Calculations are made for the complete linear stability equations
in cylindrical coordinates. Several classes of the exact solutions
of the Navier-Stokes equations describing spatially developing and
unsteady flows, for which the linear stability problems can be
rigorously reduced to eigenvalue problems of ordinary differential
equations, are defined. Those exactly solvable nonparallel and
unsteady flow stability problems can be used for testing
approximate approaches and the methods based on direct numerical
simulations of the (linearized) Navier-Stokes equations. The exact
solutions of the viscous incompressible Navier-Stokes equations
determined as the basic states, for which the linear stability
problem is exactly separable, may be themselves of interest from
theoretical and engineering points of view.

\end{abstract}


\maketitle

\section{Introduction}

The classical linear stability theory of viscous incompressible
flows is concerned with the development in space and time of
infinitesimal perturbations around a given basic flow. Then small
disturbances are resolved into normal modes which, for a
steady-state basic flow, depend on time exponentially with a
complex exponent $\lambda$. For parallel shear basic flows,
further separation of variables in the governing stability
equations leads to a set of ordinary differential equations which,
with taking recourse to Squire's theorem and considering only 2-D
disturbances, reduces to the Orr-Sommerfeld equation. When this
equation is solved with proper boundary conditions, the problem of
linear stability of parallel flows is reduced to a 2-point
boundary (eigen) value problem.

For nonparallel basic flows, when the equations for disturbance
flow are dependent not only on the normal coordinate, the
corresponding operator does not separate unless certain terms are
ignored. The approximation, that neglects the nonparallel terms
and relates the stability characteristics to those of the
equivalent parallel flow. has been extensively used for the
boundary layer type flows to retain the great advantage of
reducing the disturbance equations to ordinary differential
equations (see, e.g., Reed and Saric, 1996). A number of weakly
nonparallel theories, which seek to account for the affects of the
flow divergence through equations that include higher-order terms
than those in the Orr-Sommerfeld equation, have been developed
(see reviews in Reed and Saric, 1996; Herbert 1997; Saric et al.
2003).

It is worth mentioning, in this context, the works on stability of
conical flows by Shtern and Hussain (1998), Shtern and Drazin
(2000) where an exact transformation, reducing the stability
problem to an ordinary differential equation despite the
non-parallelism of the basic flow, is found. However, the
advantage of such a transformation in the analysis is limited to
the particular class of steady perturbation modes. In Shtern and
Hussain (2003), the approach of Shtern and Hussain (1998) has been
applied to the time-oscillatory disturbances by using a far-field
approximation in the equations for the disturbances.

Several methods have been designed for numerical solution of the
nonparallel flow stability problems. In the context of
boundary-layer type flows, the most successful effort to-date is
the parabolic-stability-equation (PSE) approach, introduced and
recently reviewed by Herbert (1997). However, the PSE based
numerical studies are not able to accommodate the upstream
propagation of disturbances. The 'global' linear stability
analysis (see review by Theofilis 2003) was developed for
analyzing stability of the two-dimensional flows. The nonparallel
stability effects have been also investigated on the basis of
direct numerical simulations of the linearized Navier-Stokes
equations (e.g., Davies and Carpentier 2003) or the complete
Navier-Stokes equations (e.g., Fazel and Konzelmann 1990). Such
numerical simulations are not equivalent to a stability analysis,
and, in fact, have more in common with physical experiments than
stability theory. The numerical methods suffer from the problem of
boundary conditions on the 'open', inflow and outflow, boundaries
which (especially outflow boundary conditions) can lead to
spiritus effects, even when carefully implemented.

If the basic flow is non-steady, this brings about great
difficulties in theoretical studies of the instability since the
method of normal modes in its traditional form, with the modes
depending on time as $exp (\lambda t)$, is not applicable at all.
(Some success has been achieved in studying stability of the
time-periodic basic states when Floquet theory can be applied -
see Drazin and Reid 1981.) If an unsteady flow is non-parallel, it
should further complicate matters. As a matter of fact, there are
no examples of the linear stability problem for viscous
incompressible flows developing both in space and time which is
exactly solvable via separation of variables.

Recently, the so-called direct approach to separation of variables
in linear PDEs has been developed by a proper formalizing the
features of the notion of separation of variables (see, e.g.,
Zhdanov and Zhalij 1999a, 1999b). In this approach, a form of the
'ans\"atz' for a solution with separated variables in a new
coordinate system as well as a form of reduced ODEs, that should
be obtained as a result of the variable separation, are postulated
from the beginning. The method has been successfully applied to
several equations of mathematical physics (see, e.g., Zhalij,
1999; Zhdanov and Zhalij, 1999a, 1999b; Zhalij, 2002).

In the present paper, we apply this approach to the linear
stability equations which govern the disturbance behavior in
viscous incompressible fluid flows. The calculations are made for
the linear stability equations written in cylindrical coordinates.
The new coordinate systems and the forms of basic flows, which
permit the postulated form of separation of variables in the
equations for disturbances, are determined as the result of
application of the method. Then the basic flows are specified by
the requirement that they exactly satisfy the Navier-Stokes
equations.

The paper is organized as follows. In Section II, we give a
description of the method, as applied to the problem of linear
stability of a three-dimensional unsteady basic flow with respect
to the three-dimensional unsteady perturbations, and present the
results. An example of application of the method to the linear
stability equations with a restriction to the two-dimensional
perturbations is also presented. In Section III, we discuss the
fluid dynamics interpretation of some basic flows, defined in
Section II as the exact solutions of the Navier-Stokes equations
possessing exactly solvable stability problems, and formulate the
corresponding two-point boundary value problems of ordinary
differential equations. Concluding remarks on the results obtained
are furnished in Section IV.

\section{Application of the Direct Method to separation of variables in the stability equations}

\subsection{Procedure}

We formulate the linear stability problem based on the
Navier-Stokes equations written in cylindrical coordinates ($r,
\varphi, z$). As usual in stability analysis, we split the
velocity and pressure fields $(\hat v_r, \hat v_{\varphi}, \hat
v_z,\hat p)$ into two problems: the basic flow problem $(V_r,
V_{\varphi}, V_z, P)$ and a perturbation one
$(v_r,v_{\varphi},v_z, p)$,
\begin{equation}
\label{Stab1} \hat v_r=V_r+v_r,\quad \hat
v_{\varphi}=V_{\varphi}+v_{\varphi},\quad \hat v_z=V_z+v_z,\quad
\hat p=P+p
\end{equation}
Introducing (\ref{Stab1}) into the Navier-Stokes equation written
in terms of the variables $(\hat v_r, \hat v_{\varphi}, \hat
v_z,\hat p)$ and neglecting all terms that involve the square of
the perturbation amplitude while imposing the requirement that the
basic flow variables $(V_r, V_{\varphi}, V_z, P)$ themselves
satisfy the Navier-Stokes equations, one arrives at the following
set of linear stability equations in cylindrical coordinates:

\begin{eqnarray} && \frac{\p v_r}{\p t} + V_r
\frac{\p v_r}{\p r} + v_r \frac{\p V_r}{\p r} +
        \frac{V_\vp}{r} \frac{\p v_r}{\p \vp} +
        \frac{v_\vp}{r} \frac{\p V_r}{\p \vp} + V_z \frac{\p v_r}{\p z} +
        v_z \frac{\p V_r}{\p z} - 2\frac{V_\vp v_\vp}{r} = \non \\
&&
    \qquad  -\frac{1}{\rho} \frac{\p p}{\p
            r} + \nu \left(\frac{\p^2 v_r}{\p r^2} +
              \frac{1}{r^2} \frac{\p^2 v_r}{\p \vp^2} + \frac{\p^2 v_r}{\p z^2} +
              \frac{1}{r}\frac{\p v_r}{\p r} - \frac{2}{r^2} \frac{\p v_\vp}{\p \vp} -
              \frac{v_r}{r^2}\right),\non \\
&& \frac{\p v_\vp}{\p t} + V_r \frac{\p v_\vp}{\p r} +
        v_r \frac{\p V_\vp}{\p r} +
        \frac{V_\vp}{r} \frac{\p v_\vp}{\p \vp} +
        \frac{v_\vp}{r} \frac{\p V_\vp}{\p \vp} +
        V_z \frac{\p v_\vp}{\p z} + v_z \frac{\p V_\vp}{\p z} +
        \frac{V_r v_\vp}{r} +
        \frac{v_r V_\vp}{r} =\non \\
&& \qquad -\frac{1}{\rho r} \frac{\p
            p}{\p \vp} + \nu \left(\frac{\p^2 v_\vp}{\p r^2} +
              \frac{1}{r^2} \frac{\p^2 v_\vp}{\p \vp^2} +
              \frac{\p^2 v_\vp}{\p z^2} + \frac{1}{r}\frac{\p v_\vp}{\p r} +
              \frac{2}{r^2} \frac{\p v_r}{\p \vp} - \frac{v_\vp}{r^2}\right),\label{1.1} \\
&& \frac{\p v_z}{\p t} + V_r \frac{\p v_z}{\p r} + v_r \frac{\p
V_z}{\p r} +
        \frac{V_\vp}{r} \frac{\p v_z}{\p \vp} +
        \frac{v_\vp}{r} \frac{\p V_z}{\p \vp} + V_z \frac{\p v_z}{\p z} +
        v_z \frac{\p V_z}{\p z} = \non \\
&& \qquad -\frac{1}{\rho} \frac{\p p}{\p
            z} + \nu \left(\frac{\p^2 v_z}{\p r^2} +
              \frac{1}{r^2}  \frac{\p^2 v_z}{\p \vp^2} + \frac{\p^2 v_z}{\p z^2} +
              \frac{1}{r}\frac{\p v_z}{\p r}\right),\non \\
&& \frac{\p v_r}{\p r} + \frac{1}{r} \frac{\p v_\vp}{\p \vp} +
\frac{\p v_z}{\p z} +
        \frac{v_r}{r}=0,\non
\end{eqnarray}

Let us introduce a new coordinate system $t$, $\xi=\xi(t, r)$,
$\gamma=\gamma(t,\varphi)$, $\eta=\eta(t, z)$.

We will say that the system (\ref{1.1}) is {\it separable in the
non-stationary cylindrical coordinate system} $\xi, \gamma, \eta$
if the separation ans\"atz,
\begin{eqnarray}
&&v_r = T(t)\exp(a \eta + m \gamma + s S(t)) f(\xi),\non\\
&&v_\varphi = T(t)\exp(a \eta + m \gamma + s S(t)) g(\xi),\label{2.1}\\
&&v_z = T(t)\exp(a \eta + m \gamma + s S(t)) h(\xi),\non\\
&&p = T_1(t)\exp(a \eta + m \gamma + s S(t)) \pi(\xi)\non
\end{eqnarray}
reduces the system of PDEs (\ref{1.1}) to a system of three
second-order and one first order ordinary differential equations
for four functions $f(\xi), g(\xi), h(\xi), \pi(\xi)$ of the
following form
\begin{eqnarray}
&&h''(\xi)=U_{11}g'(\xi)+ U_{12}h'(\xi) + U_{13} \pi'(\xi)
+ U_{14} f(\xi) + U_{15} g(\xi)+ U_{16} h(\xi)+ U_{17} \pi(\xi),\non\\
&&f''(\xi)= U_{21}g'(\xi)+ U_{22}h'(\xi) + U_{23} \pi'(\xi)
+ U_{24} f(\xi) + U_{25} g(\xi)+ U_{26} h(\xi)+ U_{27} \pi(\xi),\label{2.2}\\
&&g''(\xi)=U_{31}g'(\xi)+ U_{32}h'(\xi) + U_{33} \pi'(\xi)
+ U_{34} f(\xi) + U_{35} g(\xi)+ U_{36} h(\xi)+ U_{37} \pi(\xi),\non\\
&&f'(\xi)=U_{41}f(\xi)+U_{42}g(\xi)+U_{43}h(\xi)+U_{44}\pi(\xi)\non.
\end{eqnarray}
Here $U_{ij}$ are second order polynomials with respect to
spectral parameters $a,s,m$ with coefficients, which are some
smooth functions on $\xi$.

Note, that equations (\ref{2.1})-(\ref{2.2}) form the input data
of the method. We can change these conditions and thereby modify
the definition of separation of variables. For instance, we can
change the order of the reduced equations (\ref{2.2}) or the
number of essential parameters $a,s,m$. So, our claim of obtaining
the {\it complete description} of basic flows and non-stationary
coordinate systems providing separation of variables in
(\ref{1.1}) makes sense only within the framework of the given
definition. If one uses a more general definition, it might be
possible to construct new coordinate systems and basic flows
providing separability of the system (\ref{1.1}).

The principal steps of the procedure of variable separation in the
system (\ref{1.1}) are as follows.
\begin{enumerate}
\item{We insert the ans\"atz (\ref{2.1}) into equation (\ref{1.1})
and express the
    derivatives $f''(\xi)$, $g''(\xi)$, $h''(\xi), f'(\xi)$,
    in terms of functions $g'(\xi)$, $h'(\xi)$, $\pi'(\xi)$,
    $f(\xi)$, $g(\xi)$, $h(\xi)$, $\pi(\xi)$ using equations (\ref{2.2}).}
\item{We regard $g'(\xi)$, $h'(\xi)$, $\pi'(\xi)$,
    $f(\xi)$, $g(\xi)$, $h(\xi)$, $\pi(\xi)$ as the new independent variables.

As the functions $\xi(t, r),  \gamma(t,\varphi), \eta(t, z), T(t),
T_1(t), S(t)$, basic flows $V_r, V_\vp, V_z$ and coefficients of
the polynomials $U_{ij}$ (which are some smooth functions on
$\xi$) are independent on these variables, we can demand that the
obtained equality is transformed into identity under arbitrary
$g'(\xi)$, $h'(\xi)$, $\pi'(\xi)$, $f(\xi)$, $g(\xi)$, $h(\xi)$,
$\pi(\xi)$. In other words, we should split the equality with
respect to these variables. After splitting we get an
overdetermined system of nonlinear partial differential equations
for unknown functions $\xi(t, r),  \gamma(t,\varphi), \eta(t, z),
T(t), T_1(t), S(t)$, basic flows $V_r, V_\vp, V_z$ and
coefficients of the polynomials $U_{ij}$.}
\item{After solving the
above system we get an exhaustive description of coordinate
systems providing separability of equations (\ref{1.1}) according
to our definition.}
\end{enumerate}

Thus, the problem of variable separation in equation (\ref{1.1})
reduces to integrating the overdetermined system of PDEs for
unknown functions $\xi(t, r),  \gamma(t,\varphi), \eta(t, z),
T(t), T_1(t), S(t)$, basic flows $V_r, V_\vp, V_z$ and
coefficients of the polynomials $U_{ij}$. This have been done with
the aid of {\em Mathematica} package.

\subsection{Results}

We will consider the stability problems with separated variables
for the basic flows specified by the requirement that they exactly
satisfy the Navier-Stokes equations.

\subsubsection{Three-dimensional perturbations}

The forms of the perturbations $v_r, v_\vp, v_z$ and $p$ are:
\begin{eqnarray}
\label{PertForm}&& v_r = T(t)\exp\left(a \eta + m \vp + s \int T(t)^2 dt\right)f(\xi),\nonumber\\
&& v_\vp = T(t)\exp\left(a \eta + m \vp + s \int T(t)^2 dt\right)g(\xi),\nonumber\\
&& v_z = T(t)\exp\left(a \eta + m \vp + s \int T(t)^2 dt\right)h(\xi),\nonumber\\
&& p = \rho T(t)^2\exp\left(a \eta + m \vp + s \int T(t)^2
dt\right)\pi(\xi).
\end{eqnarray}
where
\begin{equation}
\label{BF02} \xi = T(t)r, \qquad  \eta = T(t)z + c(t).
\end{equation}

Two classes of basic flows satisfying the Navier-Stokes equations
are found as the result of the analysis. Velocity fields for both
classes are defined by
\begin{eqnarray}
\label{Case I}
&&V_z = A(\xi)T(t)-\frac{z T'(t)}{T(t)}-\beta(t),\qquad \beta(t)=\frac{c'(t)}{T(t)}\nonumber\\
&&V_r = B(\xi)T(t)-r \frac{T'(t)}{T(t)},\nonumber\\
&& V_\vp = C(\xi)T(t),
\end{eqnarray}
where the functions $T(t)$ and $B(\xi)$ are specified in different
ways for the two classes.
\begin{equation}
\label{BF11}Class\; I:\qquad T(t)=\frac{1}{\sqrt{t}}, \qquad
B(\xi)=-\frac{3\xi}{4}+\frac{k}{\xi},
\end{equation}
where the functions $A(\xi)$ and $C(\xi)$ satisfy the equations
\begin{eqnarray}
(4k+3\xi^2-4\nu)A'(\xi)+\xi(-4k+3\xi^2+4\nu)A''(\xi)+4\xi^2\nu
A'''(\xi)=0,\nonumber \\
-4\nu k_0\xi+(-4k+3\xi^2-4\nu)C(\xi)+\xi(-4k+3\xi^2+4\nu)C'(\xi)+4\nu\xi^2 C''(\xi)=0.\label{EqnC1}
\end{eqnarray}
and the pressure distribution is given by
\begin{eqnarray}
\label{PCl1}\frac{P}{\rho}=\frac{\nu
k_0\varphi}{t}+\frac{x^2}{8t^2}
+x\left[\beta'(t)+\frac{\beta(t)}{2t}+t^{-3/2}\left(\nu A''(\xi)-\frac{4k-3\xi^2-4\nu}{4\xi}A'(\xi)\right)\right]\nonumber\\
+\frac{1}{t}\int{\frac{16k^2-5\xi^2+16\xi^2C^2(\xi)}{16\xi^3}d\xi}+p_0(t)
\end{eqnarray}

The ODEs (\ref{EqnC1}) can be explicitly solved in terms of the
incomplete gamma functions.

\medskip

\begin{equation}
\label{BF12} Class\; II:\qquad T(t)=1, \qquad
B(\xi)=\frac{k}{\xi}\nonumber
\end{equation}
where $A(\xi)$ and $C(\xi)$ satisfy the equations
\begin{eqnarray}
(k-\nu)A'(\xi)+\xi(\nu-k)A''(\xi)+\xi^2\nu A'''(\xi)=0, \nonumber
\\
\nu k_0\xi+(k+\nu)C(\xi)+\xi(k-\nu)C'(\xi)-\xi^2\nu C''(\xi)=0
\label{EqnC2}
\end{eqnarray}
and the corresponding pressure distribution is
\begin{equation}
\frac{P}{\rho}=\nu k_0\varphi+x\left(\beta'(t)+\nu A''(\xi)+\frac{\nu-k}{\xi}A'(\xi)\right)
+\int{\frac{k^2+\xi^2C^2(\xi)}{\xi^3}d\xi}+p_0(t)\nonumber
\end{equation}

The ODEs (\ref{EqnC2}) can be explicitly solved in elementary
functions.

\medskip

The equations with separated variables can be written for both
classes in the forms
\begin{eqnarray*}
&& f(\xi)(\xi^2 s + \nu - m^2 \nu - a^2 \xi^2 \nu + a \xi^2 A(\xi) + m \xi C(\xi) + \xi^2 B'(\xi))+ \\
&& \qquad 2(m \nu - \xi C(\xi)) g(\xi)+ \xi((-\nu + \xi B(\xi)) f'(\xi) + \xi(\pi'(\xi) - \nu f''(\xi))) = 0, \\
&& (\xi^2 s + \nu - m^2 \nu - a^2 \xi^2 \nu + a \xi^2 A(\xi) + \xi B(\xi) + m \xi C(\xi)) g(\xi) + \\
&& \qquad f(\xi)(-2 m \nu + \xi C(\xi) + \xi^2 C'(\xi))+ \xi(m \pi(\xi) +(-\nu + \xi B(\xi)) g'(\xi) - \xi \nu g''(\xi)) = 0, \\
&& (\xi^2 s - m^2 \nu - a^2 \xi^2 \nu + a \xi^2 A(\xi) + m \xi C(\xi)) h(\xi) + \\
&& \qquad \xi(a \xi \pi(\xi) + \xi f(\xi) A'(\xi) - \nu h'(\xi) + \xi B(\xi) h'(\xi) - \xi \nu h''(\xi)) = 0, \\
&& f(\xi) + m g(\xi) + \xi(a h(\xi) + f'(\xi)) = 0.
\end{eqnarray*}

\subsubsection{Two-dimensional perturbations}

The stability properties of a given flow may be tested by
considering perturbations of specific structures. For example, the
problem may be restricted to the two-dimensional perturbations
even though an analog of the Squire theorem cannot be proved (see,
e.g., Griffond and Casalis 2001, Joslin 1996), or the perturbation
flow field may be taken to have the same general form as the basic
state (Duck and Dry 2001). Although the stability analysis
restricted to perturbations of specific forms is not complete, it
enables one to show that the flow is susceptible to a special kind
of instability. To demonstrate that a specification of the
disturbance field may lead to new possibilities we consider the
results of application of the direct method to the linear
stability equations with a restriction to the two-dimensional
perturbations with $v_z=0$ and $v_r$ and $v_\phi$ not dependent on
$z$.

The separability analysis leads to the perturbations of the form
\begin{eqnarray}\label{A1}
&& v_r = T(t) \exp\left(m \vp + s \int T(t)^2 dt\right) f(\xi),\nonumber\\
&& v_\vp = T(t) \exp\left(m \vp + s \int T(t)^2 dt\right) g(\xi),\nonumber\\
&& v_z = 0,\nonumber\\
&& p = \rho T(t)^2 \exp\left(m \vp + s\int T(t)^2 dt\right)
\pi(\xi),\nonumber\\
&& \xi = T(t) r
\end{eqnarray}
which is a particular case of (\ref{PertForm}) for $a=0$. However,
for the perturbations of the form (\ref{A1}), the corresponding
basic flows are not restricted to those listed in Section IIB1. In
addition, the following basic flows are permitted
\begin{eqnarray}\label{A2}
V_z = -k z + \beta(t),\nonumber\\
V_r = k r/2 + q/r,\nonumber\\
V_\vp = \nu B(\xi) T(t).\nonumber
\end{eqnarray}
\begin{eqnarray}
\frac{P}{\rho}=-\frac{1}{2}k^2x^2+x\left(k \beta(t)-\beta'(t)\right)
-\frac{4q^2+k^2r^4}{8r^2}+T^2(t)\left(\nu^2\int{\frac{B^2(\xi)}{\xi}d\xi}
-\frac{1}{2}\nu k_0\varphi\right)+p_0(t)\nonumber
\end{eqnarray}
where the functions $T(t)$ and $B(\xi)$ satisfy the equations
\begin{equation}T'(t)-\frac{1}{2}\left(Q T^3(t)-k T(t)\right)=0\label{EqnT} \end{equation}
\[k_0\xi-(2q+2\nu+Q\xi^2)B(\xi)-\xi(2q-2\nu+Q\xi^2)B'(\xi)+2\nu \xi^2B''(\xi)=0 \]
which leads to the following cases
\begin{equation}\label{A3}
T(t) =  \frac{1}{\sqrt{e^{k t}+1}} \quad \left(\frac{Q}{k}=1\right), \qquad  T(t) =
    \frac{1}{\sqrt{e^{k t}-1}}  \quad \left(\frac{Q}{k}=-1\right),\nonumber
    \end{equation}
\begin{equation}
T(t) = 1  \quad \left(\frac{Q}{k}=1\right), \qquad T(t) = e^{-k t/2}  \quad \left(Q=0\right).\nonumber
    \end{equation}

\medskip

\section{Specific flow stability problems}

In this section, we will discuss the fluid dynamics interpretation
of some basic flows, defined above as the \textit{exact solutions
of the Navier-Stokes equations} for which the corresponding
\textit{stability problems are exactly separable}, and will
formulate the corresponding two-point boundary value problems.

We will consider particular cases of the class of the exact
solutions of the Navier-Stokes equations in cylindrical
coordinates identified in Section IIB1 as \textit{Class I}.  It is
possible to enrich the solution defined by the formulas
(\ref{BF02}) - (\ref{PCl1}) using the invariance of the solution
with respect to a shift of time variable. Making change of
variables $t= t'-1/b$, where $b$ is a constant, and omitting
primes in what follows, we will have the solution of the
Navier-Stokes equations in the form
\begin{eqnarray}
\label{SOL1}&& V_z=-\frac{b
z}{2(1-bt)}+\frac{F(\zeta)}{\sqrt{1-bt}}-\beta (t),\quad
V_r=\frac{1}{\sqrt{1-bt}}\left(\frac{b \zeta}{4}+\frac{k}{\zeta}\right),\quad
V_{\phi}=\frac{M(\zeta)}{\sqrt{1-bt}},\non \\
&&\zeta
=\frac{r}{\sqrt{1-bt}}.\quad \beta (t)=c'(t)\sqrt{1-bt}
\end{eqnarray}
where $b$ can be both positive and negative.

The corresponding pressure distribution is
\begin{eqnarray}
\label{SOL2}
&&\frac{P}{\rho}=\frac{b^2x^2}{8(1-b
t)^2}+x\biggl[\beta'(t)-\frac{b\beta(t)}{2(1-b t)}
+(1-b t)^{-3/2}\biggl(\nu F''(\zeta)-\frac{(4k-4\nu+3b\zeta^2)F'(\zeta)}{4\zeta}\biggr)\biggr]\nonumber \\
&&+\frac{1}{1-b t}\int{\frac{16 k^2-5b^2\zeta^4+16 \zeta^2M^2(\zeta)}{16\zeta^3}d\zeta}-\frac{k_0\nu\vp}{1-b t}+p_0(t)
\end{eqnarray}

The functions $F(\zeta)$
and $M(\zeta)$ satisfy the equations
\begin{eqnarray}
&&\label{SOL4}(4k-3b\zeta^2-4\nu)F'(\zeta)+\zeta(-4k-3b\zeta^2+4\nu)F''(\zeta)+4\zeta^2\nu
F'''(\zeta)=0,\\
&&\label{SOL5}4
k_0\nu\zeta+(-4k-3b\zeta^2-4\nu)M(\zeta)+\zeta(-4k-3b\zeta^2+4\nu)M'(\zeta)+4\zeta^2\nu
M''(\zeta)=0.
\end{eqnarray}
Equations (\ref{SOL4}) and (\ref{SOL5}) can be solved in
quadratures
\begin{eqnarray}
\label{SOL6}
&&F(\zeta)=c_1+c_2\Gamma\left(\frac{k}{2\nu},-Z(\zeta)\right)+c_3\int{e^{Z(\zeta)}\zeta^{\frac{k}{\nu}-1}
\Gamma\left(1-\frac{k}{2\nu},Z(\zeta)\right)d\zeta},\nonumber\\
&&Z(\zeta)=\frac{3b\zeta^2}{8\nu},\\
\label{SOL7}&&M(\zeta)=\frac{1}{\zeta}\biggl[c_4+c_5\Gamma\left(1+\frac{k}{2\nu},-Z(\zeta)\right)\nonumber\\
&&+\frac{k_0}{2}\left(\frac{3b}{8\nu}\right)^{\frac{k}{2\nu}}\int{e^{Z(\zeta)}\zeta^{\frac{k}{\nu}-1}
\Gamma\left(-\frac{k}{2\nu},Z(\zeta)\right)d\zeta}\biggr],
\end{eqnarray}
where $\Gamma (A,Z)$ is the incomplete Gamma function and
$c_1,\ldots,c_5$ are arbitrary constants. Note also the expression
for $F'(\zeta)$
\begin{equation}
\label{SOL6p}
F'(\zeta)=e^{Z(\zeta)}\zeta^{\frac{k}{\nu}-1}\left[-2c_2\left(-\frac{3b}{8\nu}\right)^{\frac{k}{2\nu}}
+c_3\Gamma\left(1-\frac{k}{2\nu},Z(\zeta)\right)\right].
\end{equation}

The correspondingly specified perturbations (\ref{PertForm}) take
the forms
\begin{eqnarray}
\label{PERT1}&& v_r = (1-bt)^s\exp\left(a \eta + m \vp\right)f(\zeta),\nonumber\\
&& v_\vp = (1-bt)^s\exp\left(a \eta + m \vp \right)g(\zeta),\nonumber\\
&& v_z = (1-bt)^s\exp\left(a \eta + m \vp \right)h(\zeta),\nonumber\\
&& \frac{p}{\rho} = (1-bt)^{s-1/2}\exp\left(a \eta + m \vp
\right)\pi(\zeta).
\end{eqnarray}
where
\begin{equation}
\label{PERT2}
\eta=\frac{z}{\sqrt{1-b t}}+c(t)
\end{equation}

The equations for the perturbation amplitudes are
\begin{eqnarray}
\label{PERT3}
&& \left( -4k + b\zeta^2 - 4b\zeta^2 s + 4\nu - 4m^2 \nu - 4a^2 \zeta^2 \nu
+ 4a \zeta^2 F(\zeta) + 4m \zeta M(\zeta)\right)f(\zeta)\nonumber\\
&&\qquad + \zeta\left(4k - 4\nu + 3b \zeta^2\right) f'(\zeta)- 4\nu\zeta^2f''(\zeta)\nonumber\\
&&\qquad+8\left(m \nu - \zeta M(\zeta)\right) g(\zeta)
+4\zeta^2\pi'(\zeta) = 0, \nonumber\\
&& \left(4k + b\zeta^2 - 4b\zeta^2s + 4\nu - 4m^2 \nu - 4a^2
\zeta^2 \nu
+4 a \zeta^2 F(\zeta) + 4m \zeta M(\zeta)\right) g(\zeta)\nonumber \nonumber\\
&&\qquad +\zeta\left(4k - 4\nu + 3b \zeta^2\right) g'(\zeta)- 4\nu \zeta^2g''(\zeta)\nonumber\\
&&\qquad + \left(-8 m \nu + 4\zeta M(\zeta)+ 4\zeta^2 M'(\zeta)\right)f(\zeta)
+4m\zeta\pi(\zeta) = 0, \nonumber\\
&& -2\left(b\zeta^2+2b\zeta^2 s +2 m^2 \nu + 2a^2 \zeta^2 \nu - 2a
\zeta^2 F(\zeta)
- 2 m \zeta M(\zeta)\right)h(\zeta)\nonumber \\
&&\qquad +\zeta\left(4k - 4\nu + 3b \zeta^2\right) h'(\zeta)- 4\nu \zeta^2h''(\zeta)
+4\zeta^2F'(\zeta)f(\zeta)+ 4a\zeta^2\pi(\zeta) = 0, \nonumber\\
&& f(\zeta) + m g(\zeta) + \zeta\left(a h(\zeta) +
f'(\zeta)\right) = 0.
\end{eqnarray}
This system can be reduced to a system of two third-order
equations for two functions ($f(\zeta)$ and $g(\zeta)$, for
example).

The above formulas (\ref{SOL1})-(\ref{PERT3}) remain valid if we
introduce the nondimensional variables, with the time scale
$1/\vert b\vert$ and the correspondingly defined velocity scale.
In the dimensionless equations (we will retain the same notation
for the nondimensional variables), the parameter $b$ takes one of
the two values: $b=1$ or $b=-1$, and $\nu$ is replaced by
$1/\mathrm{Re}$ where $\mathrm{Re}$ is the Reynolds number.

Note that the solution (\ref{SOL1}) - (\ref{SOL2}) for $b=1$
undergos the finite-time 'breakdown' (see, e.g., Duck and Dry
2001, Hall et al. 1992) and the 'normal mode' forms
(\ref{PERT1})-(\ref{PERT2}) are naturally adjusted to the
description of the disturbed flow as the breakdown time $t=1$ is
approached.

Next we will consider several two-point boundary value problems
corresponding different specifications of the basic flow
(\ref{SOL1})-(\ref{SOL7}).

\medskip

\begin{figure}
\includegraphics[scale=0.75]{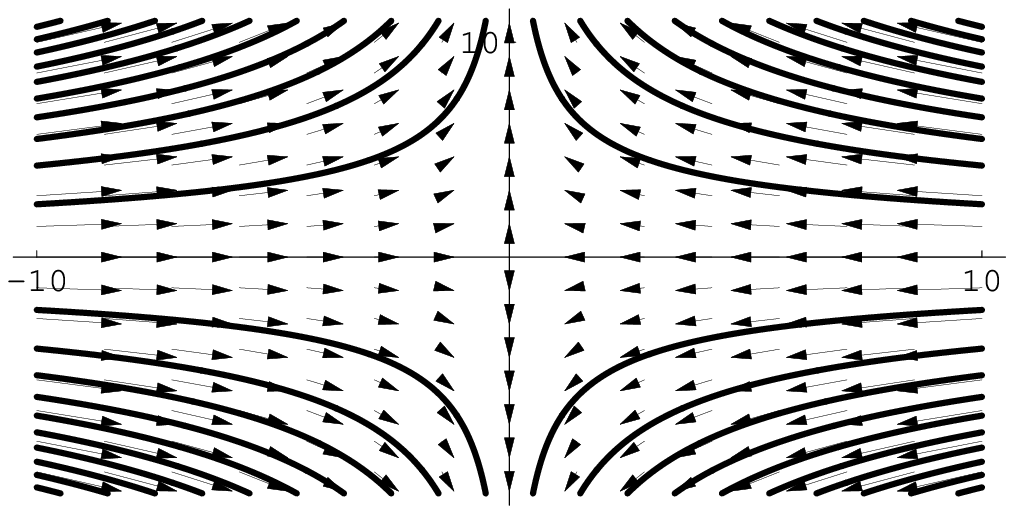}
\hspace{20pt}\includegraphics[scale=0.75]{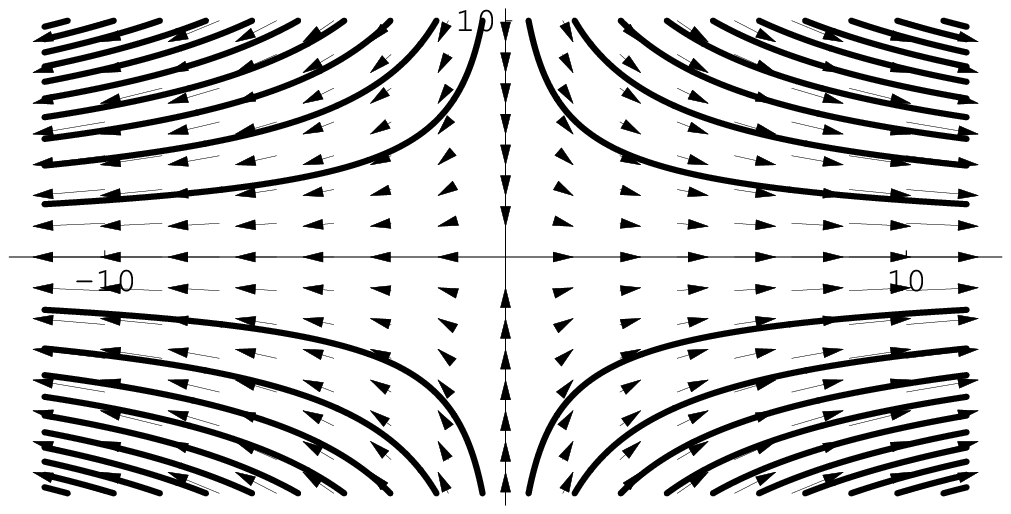}
\caption{Unsteady axially symmetric stagnation point flows: $b=1$
(left), $b=-1$ (right)}\label{StagnP}
\end{figure}

\noindent (i) \texttt{Axially symmetrical stagnation-point type
flows.} These are the simplest basic flows, that are obtained from
(\ref{SOL1})-(\ref{SOL2}) by setting
\begin{equation}
F(\zeta)=0, \; M(\zeta)=0, \; c(t)=0,\; k=0.
\end{equation}
In the case of $b=1$, the solution describes impingement of two
axially opposite stagnation point flows with velocities growing
with time, and, in the case of $b=-1$, the solution describes the
flow where fluid flowing radially from infinity approaches the
axis and spreads along it, with the flow velocity decreasing with
time as $(1+t)^{-1}$ (Fig. \ref{StagnP}). In both cases, the
boundary conditions to equations (\ref{PERT3}) are set at the axis
$\zeta=0$ and at $\zeta=\infty$, as follows
\begin{equation}
f(0)=0,\; g(0)=0,\;h'(0)=0;\;\qquad
f(\infty)=0,\;g(\infty)=0,\;h(\infty)=0
\end{equation}

\medskip

\noindent (ii) \texttt{Flow outside an expanding cylinder.} This
case corresponds to $b=-1$. The radius of the cylinder changes
with time as $R=\sqrt{1+t}$.  (In the dimensional variables,
marked with stars, it is
$R^{\ast}=R_0^{\ast}\sqrt{1-b^{\ast}t^{\ast}}$ and the value
$R_0^{\ast}$ is used as a length scale for the nondimensional
variables while the time scale is $1/\vert b^{\ast}\vert$). The
surface of the cylinder stretches in the longitudinal direction
according to the law $U=K z$ where $K=\frac{1}{2}(1+t)^{-1}$ and
$U$ is an axial velocity at $r=R$.

If the cylinder surface is impermeable, then using the boundary
condition $V_r=V$ at $r=R$, where
$V=dR/dt=\frac{1}{2}(1+t)^{-1/2}$ is the radial velocity of the
cylinder surface, yields $k=3/4$. In the case of porous cylinder.
we have $k=3/4-V_0$, where $V_0$ is a constant defining a
magnitude of the suction ($V_0>0$) or injection ($V_0<0$) velocity
as $V_s=V_0(1+t)^{-1/2}$

Since the domain is infinite in radial direction, it should be set
$c_3=0$ in the expression (\ref{SOL6p}) for $F'(\zeta)$ not to
have an unbounded behavior for $F(\zeta)$ at infinity. (The
incomplete Gamma function $\Gamma(A,Z)\sim Z^{A-1}e^{-Z}$ as
$Z\rightarrow\infty$ which results in $F'(\zeta)\sim \zeta^{-1}$
as $\zeta\rightarrow \infty$ and produces the logarithmic behavior
for $F(\zeta)$ at infinity - this can be confirmed by considering
a behavior of $F(\zeta)$ itself for specific values of $k=2n\nu$,
with $n$ being a positive integer, when closed-form solutions of
equation (\ref{SOL4}) for $F(\zeta)$ can be found.) Then the flow
at infinity represents a combination of a stagnation point flow
and a uniform stream, and the two constants in the expression for
$F(\zeta)$ are determined from the boundary condition $V_z=U$ at
$r=R$ ($F(1)=0$) and the condition for the uniform part of the
flow velocity at infinity be of the form $U_{\infty}/\sqrt{1+t}$
where $U_{\infty}$ is a constant.

\begin{figure}
\centering \includegraphics{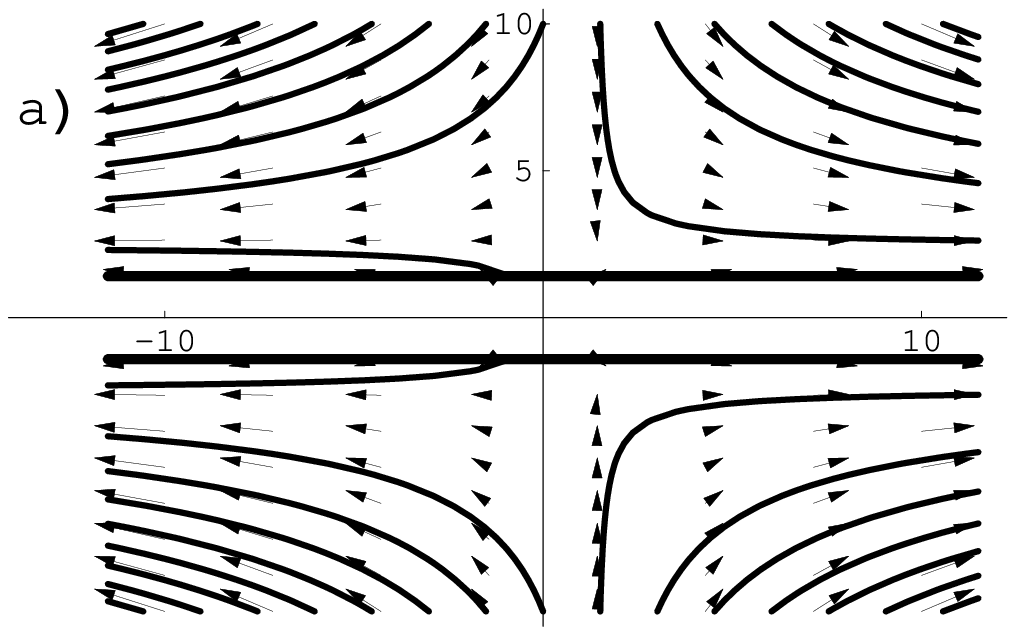}\\
\vspace{35pt} \centering \includegraphics{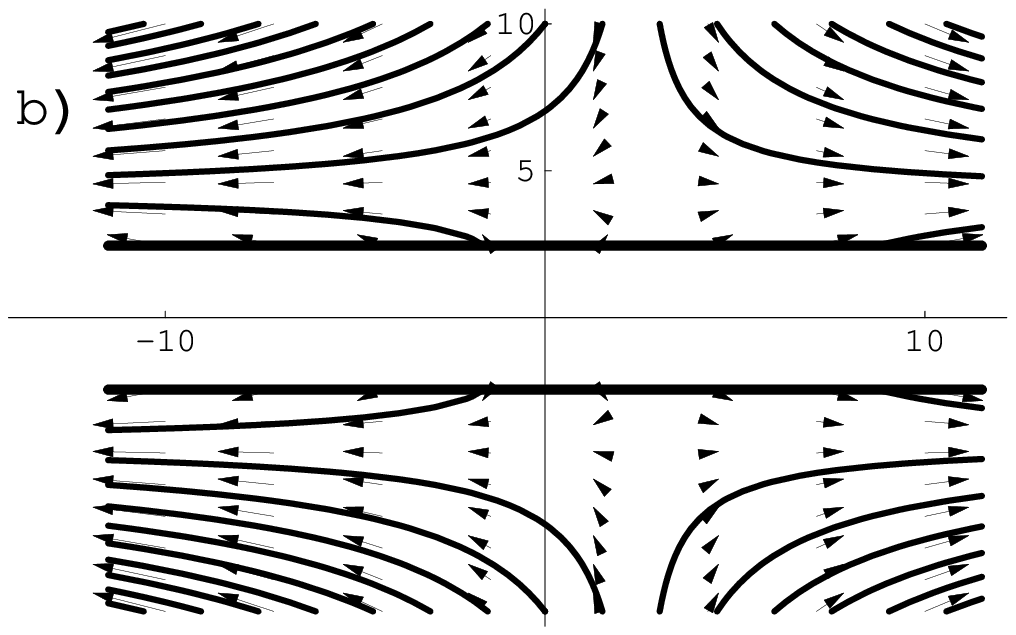}\\
\vspace{35pt} \centering
\includegraphics{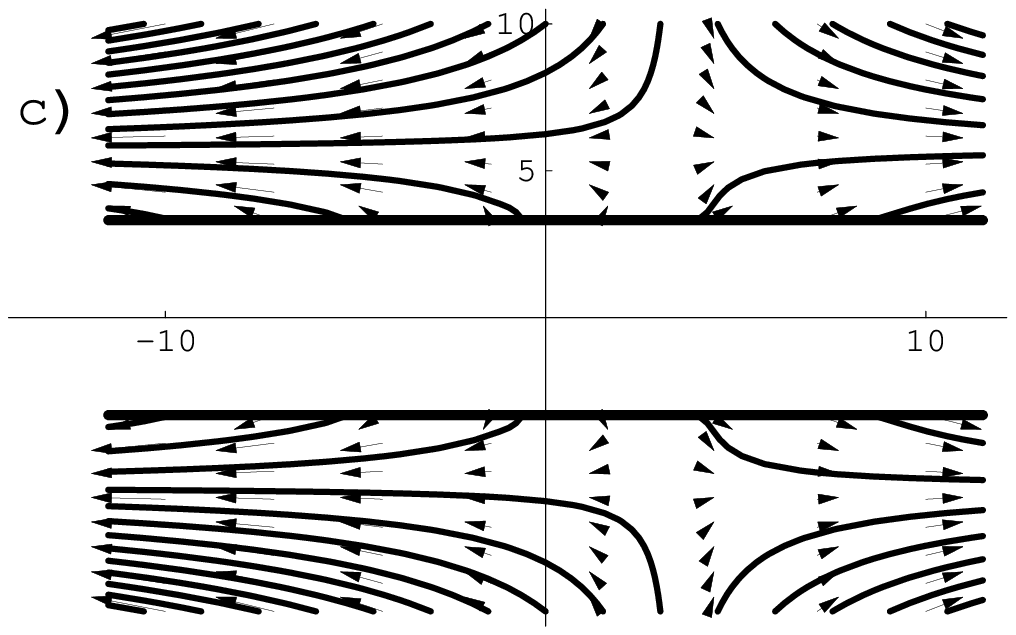} \caption{Flow outside an expanding impermeable cylinder
for $\mathrm{Re}=100$ and $U=-0.5$ at different time moments: a)
$t=1$; b) $t=5$; c) $t=10$.}\label{OutExp}
\end{figure}

For not swirling flows (a swirl can be also added with the swirl
velocity defined by (\ref{SOL1}) and (\ref{SOL7})) we have the
following to be introduced into the equations for perturbations
(\ref{PERT1})-(\ref{PERT3}):
\begin{eqnarray}
\label{Cyl1}
&&F(\zeta)=U_{\infty}\Biggl[1-\frac{\Gamma\biggl(\frac{k{\rm
Re}}{2},\frac{3{\rm Re}\zeta^2}{8}\biggr)} {\Gamma
\biggl(\frac{k{\rm Re}}{2},\frac{3{\rm
Re}}{8}\biggr)}\Biggr],\quad M(\zeta)=0,\qquad c(t)=0,\quad
k=\frac{3}{4}-V_0.\end{eqnarray} where
$\mathrm{Re}={R_0^{\ast}}^2\vert b^{\ast}\vert/\nu^{\ast}$ is the
Reynolds number (the corresponding flow structure is illustrated
by Fig. \ref{OutExp}). Note that the solution (\ref{Cyl1}) is
expressed in elementary functions for $\mathrm{Re}=2n/k$, where
$n$ is an integer, with the use of the specific values of the
incomplete Gamma function (Abramowitz and Stegun 1965):
\[\Gamma(n,Z)=1-\left(1+Z+\frac{Z^2}{2!}+\ldots
+\frac{Z^{n-1}}{(n-1)!}\right)e^{-Z}\]

In the equations for perturbations (\ref{PERT1})-(\ref{PERT3}), it
should be also set $b=-1$ and $\nu=1/\mathrm{Re}$. The boundary
conditions to equations (\ref{PERT3}) are set at $\zeta=1$ and at
$\zeta=\infty$, as follows
\begin{equation}
f(1)=0,\; g(1)=0,\;h(1)=0\; ;\qquad
f(\infty)=0,\;g(\infty)=0,\;h(\infty)=0
\end{equation}

\medskip

\noindent (iii) \texttt{Flow inside an expanding porous cylinder.}
In this case, like as in the previous one, $b=-1$, the radius of
the cylinder changes with time as $R=\sqrt{1+t}$ and the surface
stretches according to the law $U=K z$ where
$K=\frac{1}{2}(1+t)^{-1}$. The difference is in that the fluid is
now inside the cylinder and there is an injection of fluid through
the porous pipe surface, which may be either normal to the surface
or oblique, with the blowing velocity varying with time as
$V_b=V_0(1+t)^{-1/2}$, where $V_0$ is a constant.

\begin{figure}
\centering \includegraphics{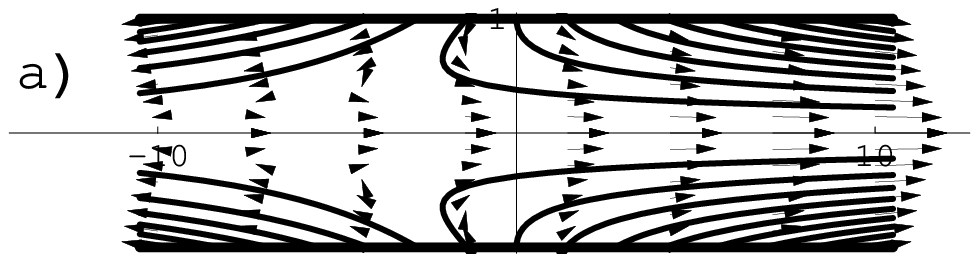}\\
\vspace{35pt} \centering \includegraphics{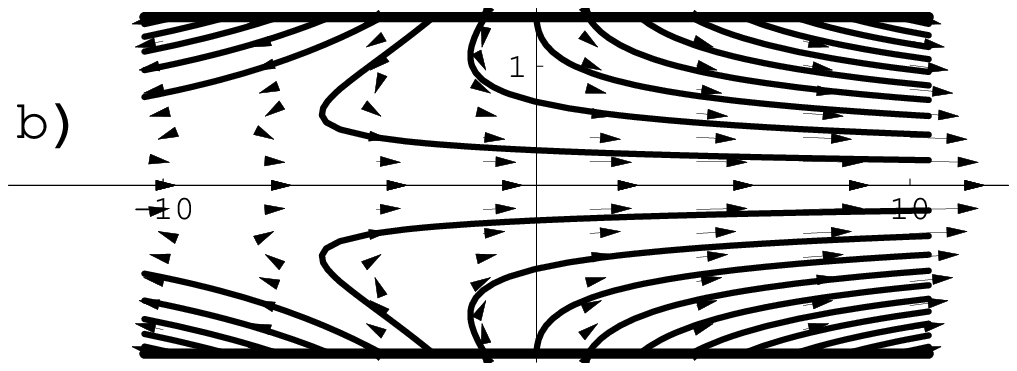}\\
\vspace{35pt} \centering
\includegraphics{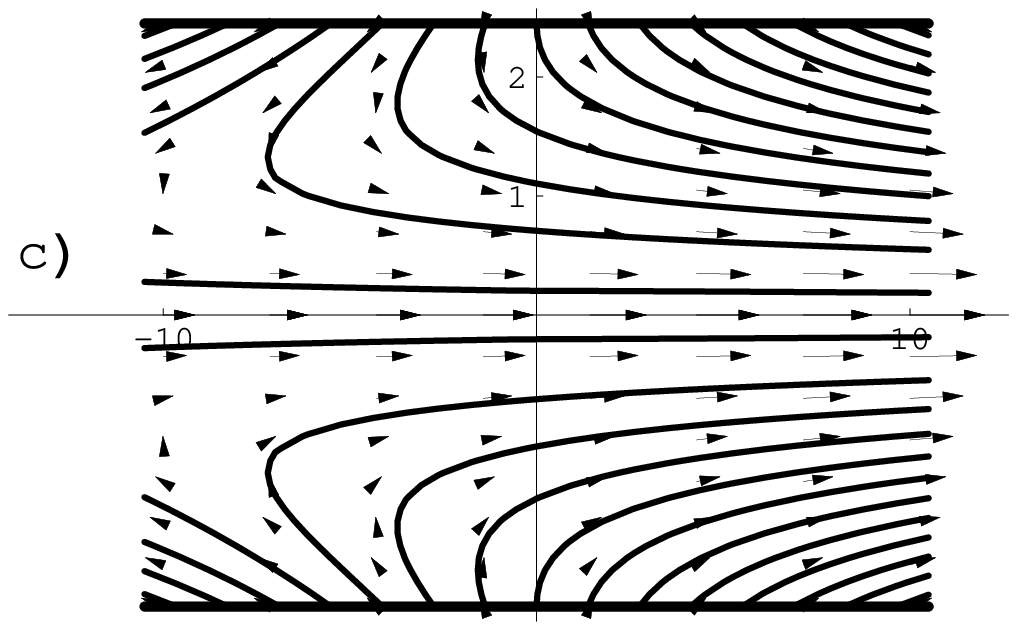} \caption{Flow inside an expanding porous cylinder
for $\mathrm{Re}=100$ and $U=5$ at different time moments: a)
$t=0$; b) $t=1$; c) $t=5$.}\label{InExp}
\end{figure}

The conditions $V_r=0$ at $r=0$ requires $k=0$ and using the
condition $\partial V_z/\partial r=0$ at $r=0$ in (\ref{SOL6p})
yields
\[F'(\zeta)= c_1\frac{1-e^{-\frac{3
\zeta^2\mathrm{Re}}{8}}}{\zeta}. \] where $\mathrm{Re}$ is the
Reynolds number. Then two arbitrary constants in the expression
for $F(\zeta)$ are determined from the condition at the cylinder
surface $F(1)=(3/4)\tanh \theta$, where the angle $\theta$ defines
the direction of blowing (with respect to the inward radial
direction), and from the condition at the axis $F(0)=U_0$, where
$U_0$ is a constant defining the axial flow velocity. Restricting
ourselves to not swirling flows and normal blowing ($\theta=0$),
we have the following to be introduced into the equations for
perturbations (\ref{PERT1})-(\ref{PERT3}):
\begin{equation}
\label{Cyl2} F(\zeta)=U_0\frac{\mathrm{Ei}\left(-\frac{3
\zeta^2\mathrm{Re}}{8}\right)-\mathrm{Ei}\left(-\frac{3
\mathrm{Re}}{8}\right)-\ln\zeta^2}{\gamma+\Gamma\left(0,\frac{3
\mathrm{Re}}{8}\right)+\ln\left(\frac{3
\mathrm{Re}}{8}\right)},\quad M(\zeta)=0,\qquad c(t)=0,\quad k=0.
\end{equation}
where $\mathrm{Ei}(Z)$ is the exponential integral function and
$\gamma$ is Euler's constant (the corresponding flow is shown in
Fig. \ref{InExp}). Note that, despite the presence of the
logarithmic term in the nominator, the expression (\ref{Cyl2}) for
$F(\zeta)$ is finite at $\zeta=0$ since the expansion of
$\mathrm{Ei}(Z)$ for small $Z$ includes the term $\ln Z$.

It should be also set $b=-1$ and $\nu=1/\mathrm{Re}$ in equations
(\ref{PERT1})-(\ref{PERT3}). The boundary conditions for the
perturbation amplitudes are set at the axis $\zeta=0$ and at the
cylinder surface $\zeta=1$, as follows
\begin{equation}
f(0)=0,\; g(0)=0,\;h'(0)=0\;;\qquad f(1)=0,\;g(1)=0,\;h(1)=0
\end{equation}

\medskip

\begin{figure}
\centering \includegraphics{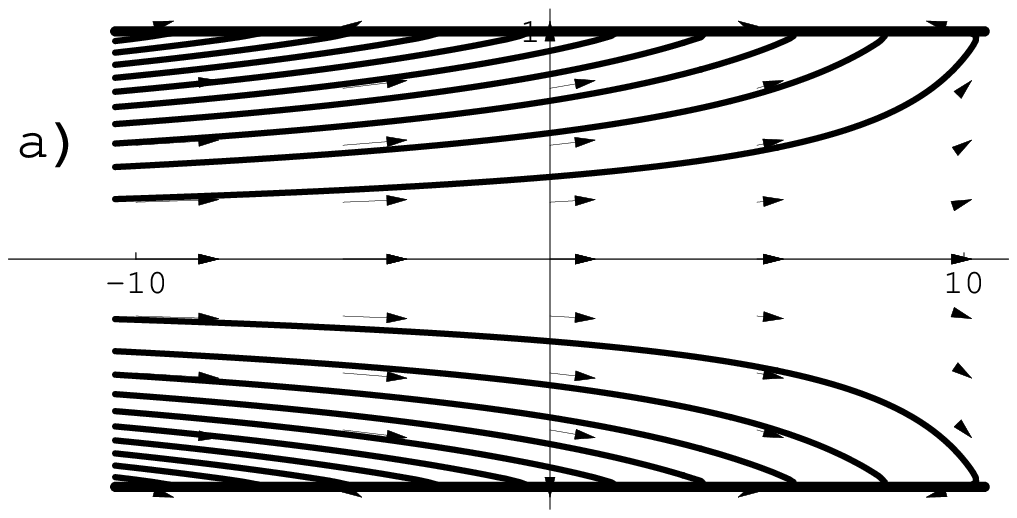}\\
\vspace{35pt} \centering \includegraphics{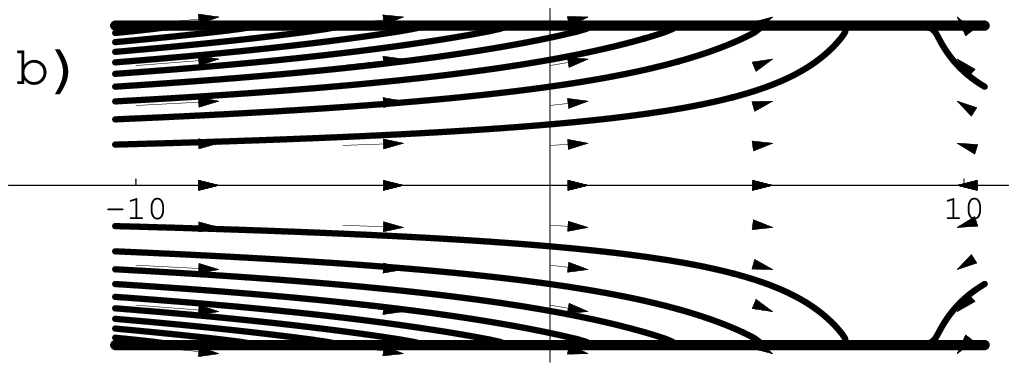}\\
\vspace{35pt} \centering
\includegraphics{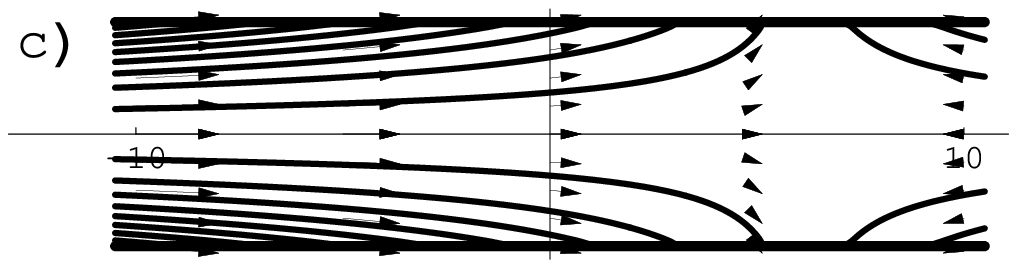} \caption{Flow inside a contracting permeable cylinder
for $\mathrm{Re}=100$ and $U=6$ at different time moments: a)
$t=0$; b) $t=0.5$; c) $t=0.75$.}\label{InContr}
\end{figure}

\noindent (iv) \texttt{Flow inside a contracting porous cylinder.}
In this case $b=1$ and the radius of the cylindrical tube changes
with time as $R=\sqrt{1-t}\; (t<1)$. The surface of the tube
shrinks according to the law $U=-K z$ where
$K=\frac{1}{2}(1-t)^{-1}$ and there is a suction of fluid through
the permeable cylinder surface, which may be either normal to the
surface ($F(1)=0$) or oblique ($F(1)=(3/4)\tanh \theta$ with the
angle $\theta$ defining the direction of suction with respect to
the outward radial direction). The suction velocity varies with
time as $V_b=V_0(1-t)^{-1/2}$ where
$V_0=\left[9/16+F^2(1)\right]^{1/2}$. For not swirling flows and
normal suction ($\theta=0$), we have the following to be
introduced into the equations for perturbations
(\ref{PERT1})-(\ref{PERT3}):
\begin{equation}
\label{Cyl3} F(\zeta)=U_0\frac{\mathrm{Ei}\left(\frac{3
\zeta^2\mathrm{Re}}{8}\right)-\mathrm{Ei}\left(\frac{3
\mathrm{Re}}{8}\right)-\ln\zeta^2}{\gamma-\mathrm{Ei}\left(\frac{3
\mathrm{Re}}{8}\right)+\ln\left(\frac{3
\mathrm{Re}}{8}\right)},\quad M(\zeta)=0,\qquad c(t)=0,\quad k=0.
\end{equation}
where $U_0$ is a constant defining the axial flow velocity (the
corresponding basic flow is shown in Fig. \ref{InContr}).

It should be set $b=1$ and $\nu=1/\mathrm{Re}$ in equations
(\ref{PERT1})-(\ref{PERT3}), and the boundary conditions are set
at the axis $\zeta=0$ and at the cylinder surface $\zeta=1$, as
follows
\begin{equation}
f(0)=0,\; g(0)=0,\;h'(0)=0\;;\qquad f(1)=0,\;g(1)=0,\;h(1)=0
\end{equation}

\medskip

\noindent (v) \texttt{Flow in the gap between concentric
cylinders.} Here, like as in the previous cases, different
boundary conditions can be considered.

\medskip

\section{Concluding remarks}

Several classes of the exact solutions of the Navier-Stokes
equations describing spatially developing and unsteady flows, for
which the linear stability problems can be rigorously reduced to
eigenvalue problems of ordinary differential equations, have been
defined. Those exactly solvable nonparallel and unsteady flow
stability problems can provide a necessary foundation for a number
approximate approaches used in the stability analysis so far. The
results can be also used for testing the methods based on direct
numerical simulations of the (linearized) Navier-Stokes equations.
Note that the basic flows considered in the paper belong to the
category of the so-called 'open' flows (see, e.g., Huerre and
Monkewitz 1990), for which the numerical instability simulations
can be quite challenging because of the problem of boundary
conditions on the inflow and outflow boundaries.

It is worth remarking that the general forms of the basic flows,
which have been obtained from the only requirement of separability
of the corresponding stability problem, are reacher than those
remaining after specification to the exact solutions of the
Navier-Stokes equations. Thus, using the approach accepted in many
stability studies, where the form of the basic flow is chosen
quite freely to approximate the physical situation of interest, we
could considerably enrich the list of relevant flows. However, our
purpose was to provide examples of a completely rigorous analysis
that reduced the stability problem to an eigenvalue problem of
ordinary differential equations.

In addition, note that we have not yet exhausted the "direct
approach" to separation of variables in the hydrodynamic stability
equations. Changing the input data of the method given by
equations (\ref{2.1})-(\ref{2.2}) may lead to new results. We
considered here only the most natural generalization of the normal
modes of the steady-state parallel flow analysis, which allows
periodicity of perturbations in two new variables, and the order
of the reduced equations was taken the same as that obtained in
the parallel flow stability problem. If one uses other input data,
it might be possible to construct new coordinate systems and basic
flows providing separability of the system (\ref{1.1}).

\medskip

We will also remark on the practical importance of specific basic
flows that have been defined in the course of our analysis (some
of them are discussed in the previous section). Those flows,
mainly, are either ones over the stretching surfaces or the flows
within porous channels possessing moving walls. The description of
the flow near a stretching surface has many important applications
in manufacturing processes in industry. A literature on the
subject (see, e.g., the book by Pop and Ingham 2001) shows
considerable research activities in this area. Solutions for
physical situations, close to those considered in our Section III,
can be found, for example, in Burde (1995a, 1995b), Youssef
(1997),  Nahapatra and Gupta (2003), Nazar et al. (2004).

Laminar, incompressible and time-dependent flows that develop
within a channel possessing permeable, moving walls have received
considerable attention in the past due to their relevance in a
number of engineering applications. Instances of direct
application of such flows include the modeling of sweat cooling or
heating, isotope separation, filtration, paper manufacturing,
irrigation, and the grain regression during solid propellant
combustion. From a different perspective, the sequences of
expansions and contractions completed by channel walls enable a
researcher to mimic more realistically peristaltic motion caused
by pulsating walls and involving fluid absorbtion and filtration
processes. For the cases, similar to those considered in the
present paper, which pertain to a pipe that exhibits either
injection or suction across porous boundaries while undergoing
uniform expansion or contraction see, e.g., Uchida and Aoki
(1977), Goto and Uchida (1990), Majdalani and Zhou (2003),
Dauenhauer and Majdalani (2003).

Thus, the exact solutions of the viscous incompressible
Navier-Stokes equations determined in this paper as the basic
states, for which the linear stability problem is exactly
separable, may be themselves of interest from both theoretical and
engineering points of view.

\medskip

\textit{Acknowledgements}. This research was supported by the
Israel Science Foundation (grant No. 117/03).

\end{document}